\documentclass[12pt]{article}
\usepackage{amssymb,amsmath}
\usepackage{latexsym}
\usepackage{amsfonts}
\usepackage{graphicx}

\def\ut#1{#1\llap{\lower2ex\hbox{$\widetilde{\hphantom{#1}}$}}}
\begin{document}

\title{Light Propagation on Quantum Curved Spacetime and Back reaction
effects}
\author{Carlos Kozameh and Florencia Parisi\thanks{%
kozameh@famaf.unc.edu.ar, fparisi@famaf.unc.edu.ar} \\
{\footnotesize {\ \textit{FaMAF, Universidad Nacional de C\'{o}rdoba, 5000 C%
\'{o}rdoba, Argentina}}}}
\date{February 26, 2007}
\maketitle

\begin{abstract}
We study the electromagnetic field equations on an arbitrary quantum
curved background in the semiclassical approximation of Loop Quantum
Gravity. The effective interaction hamiltonian for the Maxwell and
gravitational fields is obtained and the corresponding field
equations, which can be expressed as a modified wave equation for
the Maxwell potential, are derived. We use these results to analyze
electromagnetic wave propagation on a quantum Robertson-Walker space
time and show that Lorentz Invariance is not preserved. The
formalism developed can be applied to the case where back reaction
effects on the metric due to the electromagnetic field are taken
into account, leading to non covariant field equations.
\end{abstract}


\section{Introduction}

There are several claims in the literature that one of the milestones of
physics in the past century, namely Lorentz invariance, will no longer be
true once a reliable quantum theory of gravity is achieved.

In its most simple and geometrical form, this invariance can be realized by
introducing the Minkowski metric $\eta _{\mu \nu }$ as the \textquotedblleft
fixed backstage\textquotedblright\ to define distances or norms in special
relativity. For example a photon is a particle whose 4-velocity $k^{\mu }$
has zero norm, i.e.%
\begin{equation*}
\eta _{\mu \nu }k^{\mu }k^{\nu }=0,
\end{equation*}%
and the above equation is the geometrical casting of the invariance of the
speed of light. What if light travels through cosmological distances? No
problem, according to GR we simply replace the flat metric for the
nontrivial metric $g_{\mu \nu }$ of the curved spacetime. Thus, Lorentz
invariance in a local frame is generalized to covariance of the equations in
an arbitrary frame. From this point of view the above equation is a scalar
with respect to any coordinate transformation.

What if the classical metric $g_{\mu \nu }$ is replaced by a quantum
operator $\widehat{g}_{\mu \nu }$? Here we face a problem, namely, the
meaning of the resulting equation. A scalar equation can be obtained by
taking an expectation value of the operator $\widehat{g}_{\mu \nu }$ with a
suitable gravity state. If $k^{\mu }$ is a null vector with respect to $%
\langle \widehat{g}_{\mu \nu }\rangle $ for a particular state then
in general it will not be a null vector with respect to a different
gravity state. However, in this case one is not really breaking
Lorentz invariance since taking expectation values with different
states is like having classical metrics with different conformal
structure and a vector is null with respect to both metrics if and
only if the conformal structures are the same.

As a matter of fact it is difficult to imagine how to obtain non covariant
observables if one is able to construct a theory with covariant field
equations for the operators and the state vectors are invariant under the
gauge transformations resulting from the diffeomorphism group.

However, to this day we do not have such a theory. Rather, the leading
candidates for a quantum theory of gravity provide models for light
propagation on a geometry constructed from a semiclassical quantum gravity
state that break Lorentz invariance \cite{A-Camelia,GaPu}. The effects
predicted by these models are within the detection limits of present
technologies but the observed data severely compromise the validity of their
results. Both loop quantum gravity and superstring theory predict a
frequency dependent speed for photons propagating on a quantum spacetime.
However, the approximately 3000 GRBs observed by BATSE and other space
instruments showed that all the photons emitted by these bursts have the
same flight time within the instruments sensitivity limits. Another
prediction of loop quantum gravity concerns the rotation of the polarization
direction for linearly polarized radiation but the observed synchrotron
radiation for sources located at cosmological distances put severe
restrictions on the phenomenological coupling constant of the model \cite%
{GleiKo,Jacobson1}. It appears very likely that Lorentz invariance is
preserved at the linearized approximation since the observed evidence points
in this direction.

\bigskip Upon a close look at the two models one finds possible resolutions
to the conundrums. For example, the dispersion relation for photons obtained
from the superstring model can be written as \cite{Ellis}%
\begin{equation*}
g_{\mu \nu }(k)k^{\mu }k^{\nu }=0,
\end{equation*}%
that is, the metric of the spacetime depends on the energy of the photon.
The equation is fully covariant and it does not say that different photons
move with different speeds. It only says that, for the specific spacetime
constructed with one photon interacting with gravity, the metric also
depends on the energy of the photon. If one changes the energy of the photon
one is also changing the spacetime and thus, it is \textquotedblleft
illegal\textquotedblright\ in GR to compare results coming from two
different spacetimes. One could try to solve the problem of two photons with
different energies interacting with a quantum gravity state but it appears
very likely that if the problem is well set both photons should follow null
geodesics as the observations suggest.\newline

The loop quantum gravity model also admits a second look that offers an
alternative explanation for the unobserved prediction. If we assume that the
classical electromagnetic field is the 2-form $F_{\mu\nu}$, Lorentz
invariance is preserved \cite{KoPa}, whereas if we assume the canonical
variables $(A_{\mu},-E^{\nu})$ can be regarded as classical objects, the
invariance is broken. The problem arises from the relationship
\begin{equation*}
E^{\mu}=g^{\mu\nu}E_{\nu}.
\end{equation*}%
When the metric becomes a quantum operator $\widehat{g}^{\mu\nu}$ and one
takes an expectation value of the above equation with a pure gravity state
then both the covariant and contravariant versions of the electric field
cannot be regarded as classical variables. If one assumes that the field of
the l.h.s. is not affected by the gravity state one obtains non-lorentzian
equations of motion. If, one the other hand, one assumes that the electric
field of the r.h.s is transparent to the action of the expectation value
then Lorentz invariance is preserved at a semiclassical approximation, i.e.,
neglecting back reaction effects. If the invariance is broken it must come
from taking into account these back reaction effects.\newline

The aim of this work is to study the propagation of light interacting with a
semiclassical quantum gravity state when back reaction effects are included
in this interaction. The goal is to see whether or not covariant equations
of motion are obtained for this propagation. A more precise or technical
meaning of this problem is given in Section 3 were it is defined and
analyzed. However, in section 2 we address the propagation of photons on a
non-flat semiclassical gravity state since these results are later used in
the main section of this work. The derivations on Section 2 can also be used
as a review of our previous work or as a toy model for photons propagating
on a geometry given by semiclassical quantum gravity states that are peaked
around a non trivial classical metric, as for example a Robertson Walker
spacetime. Finally, in the Conclusions we summarize our work. \bigskip

\section{Propagation of electromagnetic radiation on a non-flat
semiclassical geometry.}

In this section we analyze the interaction of quantum gravity and Maxwell
fields acting on quantum states that are a direct product of coherent states
for the electromagnetic field and weave states for gravity. The non-trivial
equations of motion that are obtained from such a scheme is called the
semiclassical approximation of loop quantum gravity. A derivation of these
equations follows

\subsection{The effective interaction Hamiltonian in the semiclassical
approximation}

Assume an arbitrary background metric $g_{\mu\nu}$, and consider a 3+1
splitting of the spacetime by introducing a foliation of space-like
hypersurfaces. We can set coordinates $(t,\vec{x})$ adapted to the foliation
such that the lapse and shift $N$ and $N^a$ are $1$ and $0$ respectively
(particular gauge choice). Let $q_{ab}$ be the induced 3-metric on the $%
\Sigma_t$ (corresponding to $t=const$) surface\footnote[1]{%
From now on we will use Latin indices to denote spacial components
and Greek indices for 4-dimensional components.}.

We will assume that there exists a geometric weave state $|\Delta \rangle $
on $\Sigma _{t}$ such that, given the classical metric $q_{ab}$, $\langle
\Delta |\hat{q}_{ab}|\Delta \rangle =q_{ab}+\mathcal{O}\left( \frac{\ell _{P}%
}{\Delta }\right) $, where $\hat{q}_{ab}$ is the quantum operator associated
to the metric tensor, $\ell _{P}$ is Planck's length, and $\Delta $ is the
typical length of the weave $|\Delta \rangle $. Such a state could be
constructed, for example, by introducing random oriented Planck scale
circular loops that form a graph adapted to the local geometry, and
considering the product of the traces of the holonomies along these loops
\cite{WG2}.\newline

Now, if $E^a$ and $B^a$ are the electric and magnetic (purely spacial)
fields on that background, the Hamiltonian density that couples these fields
to gravity is given by

\begin{eqnarray}
\mathcal{H}_{EB}&=&\frac{1}{2}\int_{\Sigma_t} d^{3}xq_{ab}\sqrt{det(q)}%
\left(E^aE^b+B^aB^b\right)  \notag \\
&=& \frac{1}{2}\int_{\Sigma_t} d^3x q_{ab}\llap{\lower2ex\hbox{$\widetilde{%
\hphantom{q_{ab}}}$}} (e^a e^b+ b^a b^b),
\end{eqnarray}

where in the last line we have rewritten the Hamiltonian in terms of the
vector densities $e^a$ and $b^a$ associated with $E^a$ and $B^a$, and $q_{ab}%
\llap{\lower2ex\hbox{$\widetilde{\hphantom{q_{ab}}}$}}$ is the 3-metric
divided by its determinant. When the Hamiltonian is expressed in these
variables, it is possible to implement Thiemann\'{}s regularization
procedure, which consists of a point splitting method where the operator
associated with the metric divided by its determinant is written as the
product of two operators $\hat{w}^k_a(\vec{x})$ (each one given by the
commutator of the Ashtekar connection $A_a^i$ and the square root of the
volume operator associated with $q_{ab}$, i.e. $\hat{w}^k_a(\vec{x})\equiv%
\left[A_a^i(\vec{x}),\sqrt{V(\vec{x})}\right]$ \cite{Thiemann3, Thiemann4})
evaluated at different points. In this way, the quantum operator
corresponding to the electric part of the Hamiltonian density is

\begin{equation}
\mathcal{H}_{E}=\frac{1}{2}\int d^{3}x\int d^{3}y\delta _{kl}\hat{w}_{a}^{k}(%
\vec{x})\hat{w}_{b}^{l}(\vec{y})e^{a}(\vec{x})e^{b}(\vec{y})f_{\epsilon }(%
\vec{x}-\vec{y}),  \label{hamiltonian}
\end{equation}

and similar for the magnetic part. Here, $f_{\epsilon}(x-y)$ is a
regularization function that tends to $\delta(x-y)$ as $\epsilon\rightarrow0$%
. The next step in the regularization procedure consists of introducing a
triangulation of the hypersurface $\Sigma_t$ into tetrahedra adapted to the
graph associated with the weave state $|\Delta\rangle$ considered \cite%
{Thiemann3,Thiemann4}.

The effective interaction Hamiltonian is then defined as the expectation
value of the above operator in a semiclassical state that is given by the
weave described before for the gravitational sector, and we will assume, in
addition, that this state is close to a coherent state for the Maxwell
sector, in such a way that, within our approximation, we can consider the
electromagnetic field as a classical quantity. Under these assumptions, the
effective Hamiltonian is given by

\begin{eqnarray}
\mathcal{H}_{E}&=&\frac{1}{2}\int d^3x \int d^3y \delta_{kl}
\langle\Delta|\hat w_a^k(\vec{x})\hat w_b^l(\vec{y})|\Delta\rangle e^a(\vec{x%
})e^b(\vec{y})f_{\epsilon}(\vec{x}-\vec{y})  \notag \\
&=& \frac{1}{2}\sum_{v_i,v_j} \delta_{kl}\langle\Delta|\hat w_a^k(v_i)\hat
w_b^l(v_j)|\Delta\rangle e^a(v_i)e^b(v_j),
\end{eqnarray}

since, by construction, the operators $\hat w_a^k(\vec{x})$ only act at the
vertices $v_i$ of the graph.\newline

If we now assume that the variation scale of $e^a$ is large compared to the
typical length of the weave state, we can expand it in a Taylor series
around the central point $\vec{x}$ of the graph. Keeping only linear terms
in $\ell_P$, and using the fact that the Hamiltonian is invariant under
spacial rotations, we find that the electric part of the effective
Hamiltonian is given, up to linear order, by:

\begin{equation}
\mathcal{H}_{Eeff}=\sqrt{det(q)}\left( \frac{1}{2}q^{ab}E_{a}E_{b}+\xi \ell
_{P}\sqrt{det(q)}e^{abc}E_{a}\nabla _{b}E_{c}\right) ,
\end{equation}

and similar for $B_a$. Here $\xi$ is a phenomenological coupling constant, $%
e^{abc}$ is the total antisymmetric tensor that represents the volume
element of the classic 3-metric $q_{ab}$, $\nabla_b$ is the 3-dimensional
covariant derivative consistent with $q_{ab}$ and we have expressed the
Hamiltonian in the original variables (instead of vector densities). Also,
we used the form $E_a$ and not the vector $E^a$ for reasons that will become
clear in the following section.

\subsection{The field equations}

To derive the field equations from the above Hamiltonian we first need to
determine the relationship between the electric and magnetic fields and the
canonical variables. Let $A$ be the Maxwell connection and $F=dA$ the
associated Maxwell 2-form. Then, the electric and magnetic fields that an
observer with 4-velocity $t^{\mu}$ would measure are given by

\begin{eqnarray}
E_{\mu}&=&F_{\mu\nu}t^{\nu}, \label{deff_E}\\
B_{\mu}&=&-\frac{1}{2}\;e_{\mu\nu}{}^{\rho\delta}F_{\rho\delta}t^{\nu},\label{deff_B}
\end{eqnarray}

respectively, where $e_{\mu\nu\rho\delta}$ is the totally antisymmetric
tensor associated to the volume element of $g_{\mu\nu}$ .

Note that, form the above expressions, both $E_{\mu}$ and $B_{\mu}$
are (as stated before) purely spacial vectors and that the
definition (\ref{deff_E}) says that $E_a=-\partial_t A_a$ (we have
chosen $A_0=0$ since we are only interested in wave propagation),
which is independent of any background metric. On the other hand, if
$\pi^a$ is the canonical momentum conjugated to $A_a$, Hamilton
equations are
\begin{eqnarray}  \label{H_2}
\partial_t A_a&=&\frac{\partial H}{\partial \pi^a}, \\
\partial_t \pi^a&=&-\frac{\partial H}{\partial A_a}.
\end{eqnarray}

These expressions are consistent with (\ref{deff_E}) only if

\begin{equation}
E_a=-\frac{\partial H}{\partial \pi^a}=\frac{\partial H}{\partial E_b} \frac{%
\partial E_b}{\partial \pi^a},
\end{equation}

which can be seen as a differential equation for $E_a(\pi^b)$. By solving it
we obtain the following relation between the electric field and the
canonical momentum

\begin{equation}
\pi^a=-\sqrt{det(q)}\left(q^{ab}E_b+2\xi \ell_P\sqrt{det(q)}
\;e^{abc}\;\nabla_c E_b\right),
\end{equation}

or, symbolically

\begin{equation}
\pi ^{a}=-H^{ab}E_{b},  \label{pi}
\end{equation}

where the \textquotedblleft metric\textquotedblright\ operator $H^{ab}$ is
given by
\begin{equation}
H^{ab}=\sqrt{det(q)}\left( q^{ab}+2\xi \ell _{P}\sqrt{det(q)}%
\;e^{abc}\;\nabla _{c}\right) .
\end{equation}

Note that eq. (\ref{pi}) is the semiclassical analogue of the relationship $%
\pi ^{a}=-\sqrt{det(q)}q^{ab}E_{b}$.

Now, the second Hamilton equation, (\ref{H_2}), leads to

\begin{equation}  \label{field_eq}
\partial_t(H^{ab}E_b)=H^{ab}e_b{}^{cd}\;\nabla_cB_d.
\end{equation}

Introducing the definitions (\ref{deff_E}) and (\ref{deff_B}) of $E_{a}$ and
$B_{a}$ in terms of the Maxwell potential we obtain the following expression

\begin{equation}
\nabla ^{\mu }F_{\mu a}=2\xi \ell _{P}\partial
_{t}[det(q)^{2}]H_{ad}e^{bcd}\nabla _{b}(F_{0c}),  \label{general_wave}
\end{equation}

where $\nabla ^{\mu }$ is the 4-dimensional covariant derivative associated
with the background metric $g$ and $H_{ab}$\ is the inverse of $H^{ab}$.
From this equation it becomes clear that the term on the left is the spatial
component of a covariant expression and the term that breaks covariance
appears in the right hand side. In particular, notice that for a stationary
metric this term vanishes and we obtain a Lorentz invariant propagation.

\subsection{Light propagation on a quantum flat FRW background}

We can apply, as an example, the formalism developed in the previous section
to the particular case of a flat Friedman-Robertson-Walker spacetime:

\begin{equation}
ds^2=-dt^2+a(t)^2(dx^2+dy^2+dz^2).
\end{equation}

In this case, the metric operator acting on a vector $C_a$ is given by

\begin{equation}
H^{ab}C_{b}=\left( a(t)\delta ^{ab}+2\xi \ell _{P}a(t)^{3}\epsilon
^{abc}\partial _{c}\right) C_{b}
\end{equation}

with $\epsilon_{abc}$ the Levi-Civita symbol, and the corresponding field
equation (\ref{field_eq}) in the semiclassical approximation is

\begin{equation}  \label{Field_eq_FRW}
\partial_t(a(t)\vec{\hat{E}})=\nabla\times\vec{\hat{B}},
\end{equation}

where we have adopted, for simplicity, vectorial notation and, for any given
vector $\vec{C}$, the quantity $\vec{\hat{C}}$ is defined as:
\begin{equation}
\vec{\hat{C}}=\vec{C}-2\xi \ell _{P}a(t)^{2}\nabla \times \vec{C},
\end{equation}%
with $\nabla \times \vec{C}$ the usual 3-D curl in flat coordinates, i.e. $%
\left( \nabla \times \vec{C}\right) _{a}\equiv \epsilon _{a}{}^{bc}\partial
_{b}C_{c}$. This is the only non trivial equation, since the other Hamilton
equation (\ref{H_1}) gives no new information, it is just the definition of
the electric field in terms of the potential. Note that in the above
expression the indices are raised and lowered with the $\delta _{ab}$
3-metric, and all the time dependence has been put explicitly.\newline

On the other hand, the classical counterpart of eq (\ref{Field_eq_FRW}) is

\begin{equation}
\partial_t(a(t)\vec{E})=\nabla\times\vec{B}.
\end{equation}

There is another equation that relates $\vec{E}$ and $\vec{B}$, and it comes
form the fact that, since $F=dA$, $dF=0$, which written in terms of the
electric and magnetic fields defined by (\ref{deff_E}) and (\ref{deff_B})
reads:

\begin{equation}
\partial_t \vec{B} = -\frac{\dot{a}}{a}\vec{B}-\frac{\nabla\times\vec{E}}{a}
\end{equation}

where $\dot{a}\equiv \frac{da}{dt}$. This expression holds both classically
and in the semiclassical approximation we are analyzing.\newline

Moreover, from the field equation (\ref{Field_eq_FRW}) we can derive the
wave equation satisfied by the Maxwell potential:

\begin{equation}  \label{wave}
\partial_t(a \partial_t \vec{A})-\frac{\nabla^2\vec{A}}{a}= 2\xi
\ell_P\left(\partial_t(a^3\partial_t(\nabla\times\vec{A}))- a
\nabla\times(\nabla^2 \vec{A})\right),
\end{equation}

while the classical covariant version is given by

\begin{equation}
\partial_t(a \partial_t \vec{A})-\frac{\nabla^2\vec{A}}{a}=0.
\end{equation}

\subsubsection{Plane Wave solutions}

Consider, as a first step, light coming from a sufficiently close source. In
that case we can take $a(t)\simeq const$ and the wave equation (\ref{wave})
reduces to

\begin{equation}
a \partial_t^2 \vec{A}-\frac{\nabla^2\vec{A}}{a}= 2\xi \ell_P a^2
\nabla\times\left(a\partial_t^2\vec{A}- \frac{\nabla^2 \vec{A}}{a}\right).
\end{equation}

If we propose a plane wave solution of the form $\vec{A}=Re\left(\vec{A}%
_0e^{iS}\right)$ and introduce it in the above expression, we obtain the
following dispersion relation for the wave vector $k_{\mu}=S_{,\;\mu}$

\begin{equation}
(-a^2k_t^2+k^2)(1-2\xi \ell_P a^2 k)=0,
\end{equation}
where $k^2 \equiv k_x^2+k_y^2+k_z^2$. We see, then, that the wave vector $%
k_{\mu}$ satisfies the usual dispersion relation $g^{\mu
\nu}k_{\mu}k_{\nu}=0 $.

Note that this result holds in the geometric approximation of wave
propagation, since in that case we are dealing with the high frequency limit
and we can neglect time derivatives of the expansion factor compared to time
derivatives of the electromagnetic field.\newline

If, on the other hand, we consider a source located at a cosmological
distance, then we can not take $a(t)\simeq const$, we must take into account
the terms containing time derivatives of $a$ and solve the complete wave
equation (\ref{wave}) perturbatively since it does nos admit plane wave
solutions with constant amplitude of the form proposed above. This is done
more easily if we introduce the conformal time $\eta $, such that $\frac{%
d\eta }{dt}=a^{-1}$. Expressed in this conformal time, the FRW line element
reads

\begin{equation}  \label{conformal_FRW}
ds^2=a(\eta)^2(-d\eta^2+dx^2+dy^2+dz^2),
\end{equation}

and the wave equation (\ref{wave}) can be rewritten as

\begin{equation}
(1-2\xi \ell _{P}a^{2}\nabla \times )\square \vec{A}=4\xi \ell
_{P}aa^{\prime }\nabla \times \vec{A}^{\prime },  \label{wave2}
\end{equation}%
\newline
where $\square \vec{A}\equiv \partial _{\eta }^{2}\vec{A}-\nabla ^{2}\vec{A}$
and prime denotes derivative with respect to the conformal time $\eta $.
Note that, in view of the conformal flatness of the metric (\ref%
{conformal_FRW}), the associated classical wave equation is simply $\square
\vec{A}=0$.

We will try to solve eq. (\ref{wave2}) in a perturbative way, by proposing a
solution of the form

\begin{equation}  \label{solution}
\vec{A}=\vec{A}_{class}+\xi\ell_P \vec{\tilde{A}},
\end{equation}

where $\vec{A}_{class}$ is the classical plane wave solution $\vec{A}%
_{class}=Re(\vec{A}_{0}e^{i(\omega \eta -\vec{k}\cdot \vec{x})})$, with $%
\omega ^{2}=k^{2}$. Introducing the solution (\ref{solution}) into (\ref%
{wave2}) and dropping terms of order $(\xi \ell _{P})^{2}$ we obtain the
equation for $\vec{\tilde{A}}$:
\begin{eqnarray}
\square \vec{\tilde{A}} &=&4aa^{\prime }\nabla \times \vec{A}%
_{class}^{\prime }  \notag  \label{eq_A_tilde} \\
&=&-4iaa^{\prime }\vec{k}\times \vec{A}_{class}^{\prime }
\end{eqnarray}%
\newline
To find the final solution, we must say something about the expansion factor
$a(t)$. Consider an Einstein-De Sitter model, that is, FRW universe
dominated by matter. In this case the radius of the universe is given by $%
a(t)=\alpha t^{2/3}$, which, written in terms of the conformal time $\eta $
is

\begin{equation}
a(\eta)=\alpha\left(\frac{\alpha}{3}(\eta-\eta_0)+t_0^{1/3}\right)^2,
\end{equation}

where the $0$ subindex denotes the moment $t_0$ of emission of the
electromagnetic radiation, at which the expansion factor is assumed to have
the value $a_0$ (this means that $\alpha=a_0t_0^{-2/3}$). Inserting this
into (\ref{eq_A_tilde}) we can obtain the final solution:

\begin{eqnarray}  \label{final_solution}
\vec{A} &=&Re\left[ e^{i(\omega \eta -\vec{k}\cdot \vec{x})}\left( \vec{A}%
_{0}+\xi \ell _{P}\alpha ^{2}(\hat{n}\times \vec{A}_{0})(L(\omega
,r)-L(\omega ,r_{0}))\right) \right] , \\
&\equiv &Re\left[ e^{i(\omega \eta -\vec{k}\cdot \vec{x})}(\vec{A}_{0}+\xi
\ell _{P}\vec{\Lambda}(\eta ,\omega ))\right] ,
\end{eqnarray}

where $\hat{n}=\frac{\vec{k}}{|\vec{k}|}$ and the function $L(\omega,r)$ is
given by

\begin{equation}
L(\omega ,r)=-\frac{\alpha ^{3}r}{9\omega ^{2}}+\frac{2\alpha r^{3}}{3}%
+i\left( \frac{\alpha ^{2}r^{2}}{3\omega }-\omega r^{4}\right) ,
\end{equation}

with $r=\frac{\alpha}{3}(\eta-\eta_0)+t_0^{1/3}$ or, in terms of the
comoving time, $r=t^{1/3}$. Note that we have imposed the condition that the
solution coincides with the classical wave at the emission time $t_0$.
\newline

We see from eq. (\ref{final_solution}) that the final solution corresponds
to a plane wave with the standard dispersion relation and hence with no
modification on the propagation speed, but with a corrected amplitude vector
due to quantum gravity effects. As we will see in the following, the
interaction of the electromagnetic radiation with the quantum spacetime
induces a frequency dependent correction on the polarization direction of
the initial wave, while its amplitude is, within the linear approximation
considered, not affected.\newline

\begin{figure}[htp]
\begin{center}
\includegraphics[angle=-90,width=8cm]{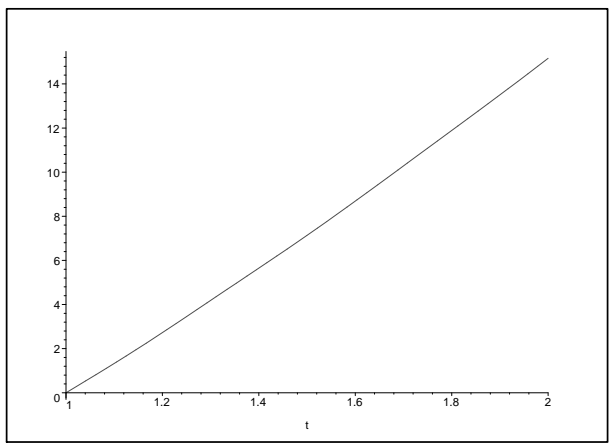} \hspace{.3cm}
\end{center}
\caption{{\protect\small Tangent of the polarization angle as a function of
the comoving time for a given photon energy, the behavior is of the form $%
\tan(\protect\theta)\propto t^{4/3}$}.}
\label{fig1}
\end{figure}

\begin{figure}[htp]
\begin{center}
\includegraphics[angle=-90,width=8cm]{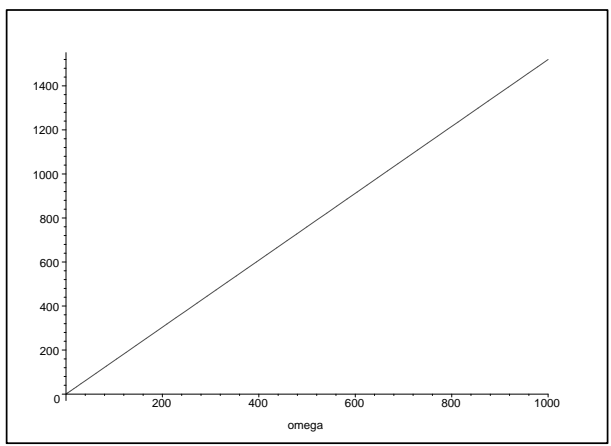} \hspace{.3cm}
\end{center}
\caption{{\protect\small Tangent of the polarization angle as a function of
the wave frequency for fixed $t$. The graph shows a linear behavior.}}
\label{fig2}
\end{figure}

To see how this correction behaves, consider a normalized and linearly
polarized initial wave. Introducing a unitary right handed basis $(\hat{e}%
_{1},\hat{e}_{2},\hat{n})$ and assuming that the initial Maxwell
potential is polarized along the $\hat{e}_{1}$ direction, we obtain,
by taking the real part of (\ref{final_solution})

\begin{eqnarray}
\vec{A}&=&[\hat{e}_1+\xi\ell_P\hat{e}_2 Re(L(\omega,r) -
L(\omega,r_0))]cos(\omega\eta-\vec{k}\cdot\vec{x})  \notag \\
&&- \xi\ell_P\hat{e}_1\alpha^2 Im(L(\omega,r) - L(\omega,r_0))sin(\omega\eta-%
\vec{k}\cdot\vec{x}).
\end{eqnarray}
\newline
From this expression we can not conclude any quantitative results
for the correction to the magnitude of the amplitude vector, since
it is given by

\begin{equation}
|\vec{A}_0+\xi \ell_P\vec{\Lambda}(\eta,\omega)|=|\vec{A}_0|+ \mathcal{O}%
((\xi\ell_P)^2)
\end{equation}

and we have been dealing with the linear approximation only (and hence
dropping quadratic terms in the whole calculation that lead to this
equation). On the other hand, the angle of rotation $\theta$ of the
polarization vector can be obtained, leading to
\begin{equation}  \label{theta}
\tan (\theta )=\xi \ell _{P}a_{0}t_{0}^{-2/3}\sqrt{Im(L(\omega ,r)-L(\omega
,r_{0}))^{2}+Re(L(\omega ,r)-L(\omega ,r_{0}))^{2}}.
\end{equation}
\newline
Here $\theta $ is measured from the initial polarization direction, and it
is an increasing function of both $t$ and $\omega $. More precisely, from (%
\ref{theta}) it is possible to prove that the tangent of the polarization
angle has, for a given frequency, a behavior where the dominant term is of
the form $t^{4/3}$ (see fig \ref{fig1}), while for a fixed instant of time
the angle grows linearly with the photon energy (fig \ref{fig2}). Notice
that this result is different from the one obtained by Gambini and Pullin
\cite{GaPu}, where the dependence of the polarization angle was quadratic in
the photon energy. Note, in addition, that the solution (\ref{final_solution}%
) does not show the birefringence effect predicted in \cite{GaPu}, since in
our formalism the propagation velocity does not depend on the frequency, nor
on the polarization state of the wave.

\vspace{.3in}

\vspace{.00000001in}

\newpage From the above considerations, we conclude that, through its flight
time, the polarization direction of an electromagnetic wave will be rotating
a frequency dependent angle $\theta (\omega )$. If a source emits a wave
packet with a continuum spectrum, the high frequency photons will rotate a
larger angle than the less energetic ones, the net result being a loss of
linear polarization. The fact that we observe, nevertheless, light coming
from cosmological sources with a high level of linear polarization is
indicative that the effect, if present, is very small (some orders of
magnitude below the sensitivity of the current instruments). Even more, we
could use the available observational data to put a bound on the coupling
constant $\xi $. Clearly, that bound would differ form the value obtained in
\cite{GleiKo,Jacobson1}, since there a quadratic effect was assumed.\newline

Just for completeness, suppose two photons are emitted simultaneously by a
cosmological source located at distance $L$ with identical (linear)
polarization state, and with wavelengths $\lambda_1$ and $\lambda_2$. Then,
at the detection time their respective polarization directions would rotate
in such a way that the difference between the corresponding angles would be
given by\vspace{.1in}
\begin{equation}  \label{relative_angle}
\Delta \theta=\xi_{rw}\ell_P \alpha\ (cL)^{4/3}\frac{c}{2\pi}\left(\frac{1}{%
\lambda_1}-\frac{1}{\lambda_1}\right).
\end{equation}
\vspace{.01in}

To derive this expression we used (\ref{theta}) and, based on observational
arguments, assumed $\theta\ll 1$ and therefore $\tan (\theta)\simeq \theta$.
We have also considered only the leading terms in (\ref{theta}) since we are
dealing with high frequencies and large distances. Hence, by using (\ref%
{relative_angle}) we could, if we knew the constant $\alpha$, obtain the
desired bound for $\xi$. However, we do not have enough information on the
parameters of the universe to determine $\alpha$, which makes it extremely
difficult to estimate that bound. On the other hand, for the curvature
effects to be appreciable, we should have observations of sources at very
high redshift, and we believe that, for the available data, the flat
background approximation suffices and the bound obtained in \cite%
{GleiKo,Jacobson1} is the most reliable.\newline

\vspace{.1in}

\textbf{Remarks:}

\begin{itemize}
\item One thing worth mentioning is that, for non-stationary geometries, the
assumption that it is possible to construct a weave state that is peaked at
that specific metric for all times is a very strong one. If the 3-metric
has, indeed, a non-trivial evolution, it is encoded in its conjugated
variable (directly related to the extrinsic curvature of the hypersurface)
and, since the weave does not satisfy the properties of a coherent sate
(namely, that it approximates the configuration variable \textbf{and} its
canonical conjugate), when one takes expectation value there is no control
on $\partial _{t}\langle \hat{q}_{ab}\rangle $ and it could happen that $%
\langle \hat{q}_{ab}\rangle $ deviates significantly form the classical
value after a short time.

\item Even more, there are indications that for arbitrary curved spaces, the
weave states might not be solutions of the Hamiltonian Constraint \cite{WG2}.%
$\ $Thus, if the weave states cannot be considered physical, the
results showed in this section may not reflect the real propagation
of light on a semiclassical FRW spacetime, assuming of course one
can find a suitable definition of semiclassical states.\newline

\item Note also that, in the derivation of the above equations, we have not
taken into account the back reaction effects on the background geometry. It
could happen that, even if the resulting equations locally preserve Lorentz
invariance (for example if one only considers stationary backgrounds), this
invariance could be broken if one considers the back reaction effect. This
possibility is analyzed in the next section.
\end{itemize}

\section{Lorentz Invariance and back reaction effects}

In the previous sections we have considered wave propagation on a fixed
background geometry, that is, we have neglected back reaction effects on the
metric due to the electromagnetic field. However, Einstein's equations
couple gravity to any other forms of energy, in particular, with the Maxwell
field, and we expect, therefore, that not only the quantum geometry will
affect the propagation of electromagnetic waves, but also that the latter
will modify, in turn, the spacetime itself. Here we will try to account for
this back reaction effect by applying the formalism developed in section 3
to the full Einstein-Maxwell theory. To do so, we first have to analyze if
it is reasonable to consider that we are within the assumptions made in that
section, namely, that we have a classical metric expressed in the
appropriate gauge, and a weave state that approximates that particular
metric.\newline

The classical equations that describe the Einstein-Maxwell theory are

\begin{eqnarray}
G_{\mu\nu}=8\pi T_{\mu\nu}, \label{Einstein}\\
\nabla^{\mu} F_{\mu\nu}=0,\label{Maxwell}
\end{eqnarray}

where the Einstein tensor $G_{\mu\nu}$ is determined by the stess-energy
tensor associated to the Maxwell field $F_{\mu\nu}$, that is

\begin{equation}  \label{stess_tensor}
T_{\mu \nu}=\frac{1}{4\pi}\left(F_{\mu \sigma}F_{\nu}{}^{\sigma} -\frac{1}{4}%
g_{\mu \nu}F_{\delta\rho}F^{\delta\rho}\right),
\end{equation}

and $\nabla ^{\mu}$ is the covariant derivative consistent with $g_{\mu\nu}$.

The idea is to solve this system of equations with an approximation scheme
were at the zeroth order the electromagnetic field propagates on a flat
background. In other words the deviation from flatness arises from the
electromagnetic stress energy tensor.

As stated above, there are two things that need to be considered:

\begin{itemize}
\item[\textit{i})] that there is a classical metric solution of (\ref%
{Einstein}) written in the gauge where $g_{tt}=-1$ and $g_{ta}=0$ (that is,
we introduce a foliation adapted to the time field $t^{\mu}$, such that the
lapse and shift functions are $N=1$ and $N^a=0$ respectively), and

\item[\textit{ii})] that it is reasonable to assume that we can construct a
semiclassical state $|\psi\rangle$ that approximates the 3-metric $q_{ab}$
induced in the spacial hypersurfaces of the above mentioned foliation, that
is, such that $\langle \psi|{\hat q}_{ab}|\psi\rangle=q_{ab}$ plus
corrections of order $\ell_P$.
\end{itemize}

If we are able to affirm these two statements, then we can assume that the
field equation for the electromagnetic field, within the semiclassical
approximation of LQG, is given by eq. (\ref{field_eq}), with $H^{ab}$ and $%
e_{abc}$ the ones corresponding to the 3-metric $q_{ab}$ associated to the
solution of (\ref{Einstein}).\newline

\textbf{The classical metric}\newline

We have to solve the field equation (\ref{Einstein}), with $T_{\mu\nu}$
given by (\ref{stess_tensor}). Since by assumption the Maxwell field
introduces a small perturbation to the Minkowski metric $\eta _{\mu\nu}$, we
will assume that the electromagnetic tensor is given by $\epsilon F_{\mu\nu}$%
, with $\epsilon $ a small parameter that will allow us to solve the
equations by means of a perturbation expansion. Then, the classical metric
will be given by

\begin{equation}  \label{pert_exp}
g_{\mu\nu}=\eta_{\mu\nu}+\epsilon g^{(1)}_{\mu\nu}+\epsilon^2
g^{(2)}_{\mu\nu}+...
\end{equation}

Besides, we will set the data (in the initial surface) of $g^{(1)}_{\mu\nu}$%
, $g^{(2)}_{\mu\nu}$, etc, equal to zero, since we are interested in the
case where there are no incoming gravitational waves, and all the
perturbation is generated by the interaction with the Maxwell field.

If we introduce (\ref{pert_exp}) into (\ref{Einstein}), and consider a
similar perturbative expansion for the electromagnetic field (and hence for
the stress-energy tensor) given by

\begin{eqnarray}
F_{\mu\nu}=\epsilon (F^{(1)}_{\mu\nu}+\epsilon F^{(2)}_{\mu\nu}+...), \\
T_{\mu\nu}=\epsilon^2 (T^{(2)}_{\mu\nu}+\epsilon T^{(3)}_{\mu\nu}+...),
\end{eqnarray}

we can solve the coupled equations order by order in a recursive way, such
that each order is determined by the previous ones. In this work we will
focus on the first non trivial corrections to the free fields.

The term of Equation (\ref{Einstein}) that corresponds to linear order in $%
\epsilon$ is simply

\begin{eqnarray}
G^{(1)}_{\mu\nu}&=&8\pi T^{(1)}_{\mu\nu} \\
&=&0,
\end{eqnarray}

since, the first non zero term of the stress-energy tensor is of order $%
\epsilon^2$, and, hence, we get the familiar result form linearized gravity
\cite{Wald}

\begin{equation}  \label{g_1}
\partial^{\delta} \partial_{(\mu}\bar{g}^{(1)}_{\nu)\delta}-\frac{1}{2}%
\;\partial^{\delta} \partial_{\delta} \bar{g}^{(1)}_{\mu \nu}-\frac{1}{2}%
\;\eta_{\mu \nu}\;\partial^{\rho}\partial^{\delta} \bar{g}^{(1)}_{\rho
\delta}=0,
\end{equation}

where $\bar{g}^{(1)}_{\mu\nu}=g^{(1)}_{\mu\nu}-\frac{1}{2}%
\eta_{\mu\nu}g^{(1)}$ with $g^{(1)}=\eta^{\mu\nu}g^{(1)}_{\mu\nu}$. Of
course the above equation reduces to the well known wave equation $%
\partial^{\delta}\partial_{\delta} \bar{g}^{(1)}_{\mu\nu}=0$ in the Lorentz
gauge. We will not, however, consider that gauge but another one consistent
with (\ref{field_eq}). We will go back to this issue later.

The only solution of eq. (\ref{g_1}) is $g^{(1)}_{\mu\nu}=0$, since, as
mentioned before, the data for this perturbation in the initial surface is
zero. On the other hand, looking at the first order in $\epsilon$ of eq (\ref%
{Maxwell}) we get

\begin{equation}
\partial^{\mu} F^{(1)}_{\mu\nu}=0,
\end{equation}

which sates, as expected, that $F^{(1)}_{\mu\nu}$ is the free Maxwell field
in flat background.

Hence, the non trivial correction for both the electromagnetic field and
metric tensor are at least of order $\epsilon^2$. The field equation for $%
g^{(2)}_{\mu\nu}$ is, in analogy to (\ref{g_1}),

\begin{equation}  \label{g_2}
\partial^{\delta} \partial_{(\mu}\bar{g}^{(2)}_{\nu)\delta}-\frac{1}{2}%
\;\partial^{\delta} \partial_{\delta} \bar{g}^{(2)}_{\mu \nu}-\frac{1}{2}%
\;\eta_{\mu \nu}\;\partial^{\rho}\partial^{\delta} \bar{g}^{(2)}_{\rho
\delta}=8\pi T^{(2)}_{\mu \nu},
\end{equation}

with

\begin{equation}
T^{(2)}_{\mu \nu}=\frac{1}{4\pi}\left(F^{(1)}_{\mu \sigma}
F^{(1)}_{\nu}{}^{\sigma}-\frac{1}{4} \eta_{\mu \nu}F^{(1)}_{\delta
\rho}F^{(1)\delta\rho}\right),
\end{equation}

and where all the indices are raised and lowered with $\eta_{\mu\nu}$. The
above expressions prove that the first correction to the metric due to back
reaction effects is generated by the free Maxwell filed. We will write, for
simplicity,

\begin{equation}  \label{corrected_metric}
g_{\mu\nu}=\eta_{\mu\nu}+\epsilon^2\gamma_{\mu\nu}
\end{equation}
\newline
with $\gamma_{\mu\nu}$ such that $\bar{\gamma}_{\mu\nu}$ is a solution of (%
\ref{g_2} ).\newline

Now, the formalism developed in the previous section requires that the
metric is written in coordinates in which it takes the form

\begin{equation}  \label{metric}
ds^2=-dt^2+q_{ab}dx^a dx^b.
\end{equation}

This can be easily done if we use the gauge freedom in eq (\ref{g_2}) that
corresponds, precisely, to a coordinate choice. It is well known that, under
the transformation $x^{\mu}\mapsto x^{\mu}+\epsilon \;\xi ^{\mu}$, the
perturbation changes according to $\gamma _{\mu\nu}\mapsto \gamma
_{\mu\nu}+\partial _{\mu}\xi _{\nu}+\partial _{\nu}\xi _{\mu}$. In order to
satisfy (\ref{metric}), the gauge choice must be such that $\gamma _{t\nu}=0$%
, which gives the four necessary conditions to fix the transformation
generator $\xi ^{\mu}$, namely

\begin{equation}
\partial_t\xi_{\nu}+\partial_{\nu}\xi_t=-\gamma_{t\nu}.
\end{equation}

Hence, it is always possible to express the metric tensor in the form (\ref%
{metric}), and, in that particular gauge, $\gamma _{\mu\nu}$ has only
spacial components.\newline

To obtain the corrections to the electromagnetic field due to back reaction
effects, we insert the solution (\ref{corrected_metric}) in the Maxwell
equation $\nabla ^{\mu}F_{\mu\nu}=0$. By doing so, we obtain

\begin{eqnarray}  \label{Maxwell_correction}
\partial^{\mu} F^{(2)}_{\mu\nu}&=&0, \\
\partial^{\mu} F^{(3)}_{\mu \nu}&=&\gamma^{\mu \sigma}\partial_{\sigma}
F^{(1)}_{\mu \nu}+ \eta^{\mu\sigma}\left(\Gamma^{(2)\delta}_{\mu
\sigma}F^{(1)}_{\delta \nu}+\Gamma^{(2)\delta}_{\nu \sigma}F^{(1)}_{\mu
\delta}\right),
\end{eqnarray}

with $\Gamma^{(2)\sigma}_{\mu\nu}$ the Christoffel symbol given by

\begin{equation}
\Gamma^{(2)\sigma}_{\mu \nu}=\frac{1}{2}\eta^{\sigma
\delta}(\partial_{\mu}\gamma_{\nu \delta}+ \partial_{\nu}\gamma_{\mu
\delta}-\partial_{\delta}\gamma_{\mu \nu}).
\end{equation}

The above equations tell us that the first non-trivial correction to the
Maxwell field is $F^{(3)}_{\mu\nu}$, since we can use an analogous argument
to the one used to conclude that $g^{(1)}_{\mu\nu}=0$; namely, we consider
that the only incoming waves are the ones given by the free Maxwell field $%
F^{(1)}_{\mu\nu}$ and hence, since there is no source to generate it, $%
F^{(2)}_{\mu\nu}$ must be zero. Therefore, the electric and magnetic fields
will be of the form

\begin{eqnarray}  \label{pert_B}
E_a&=&\epsilon E^{(1)}_a+\epsilon^3 E^{(3)}_a+..., \\
B_a&=&\epsilon B^{(1)}_a+\epsilon^3 B^{(3)}_a+....
\end{eqnarray}

In the perturbative formalism described, we saw that each order in the
perturbative expansion can be obtained from the previous ones, both for the
metric and for the electromagnetic field. We will stop the calculation here,
however, because we are just interested in the first non-trivial
corrections, which, for the Einstein-Maxwell theory, are given by eqs (\ref%
{g_2}) and (\ref{Maxwell_correction}).\newline

We have proved statement \textit{i}), namely, that it is possible to find a
classical metric solution of Einstein-Maxwell equations and write it in the
gauge where $g_{tt}=-1$ and $g_{ta}=0$. It only remains to prove the
existence of a semiclassical state that approximates that metric (statement
\textit{ii}).\newline
\newline
\newline
\newline
\textbf{The semiclassical state}\newline

We assume there exists a semiclassical state that satisfies the following
condition:

\begin{itemize}
\item It is peaked at the classical 3-metric, i.e.,
\begin{eqnarray}
\langle \psi |{\hat{q}}_{ab}|\psi \rangle &=&q_{ab\;class}+\mathcal{O}(\ell
_{P}) \\
&=&\delta _{ab}+\epsilon ^{2}\gamma _{ab}+\mathcal{O}(\ell _{P}).
\end{eqnarray}
\end{itemize}

We will derive here a formal solution of the above stated condition.\newline

In the perturbative approach we are considering, the Hamiltonian is given by
\begin{equation}
\mathcal{H}=\mathcal{H}_0+\epsilon^2\mathcal{H}_2,
\end{equation}

where $\mathcal{H}_{0}$ is the Hamiltonian constraint corresponding to pure
gravity, and $\mathcal{H}_{2}$ is the perturbation introduced by the Maxwell
field (which is the usual \textquotedblleft $q_{ab}(E^{a}E^{b}+B^{a}B^{b})$%
\textquotedblright\ term) and that describes the coupling between the two
fields. Similarly, we propose a semiclassical state of the form

\begin{equation}  \label{state}
|\psi\rangle=|\psi_0\rangle+\epsilon^2 |\psi_2\rangle,
\end{equation}

where $|\psi_0\rangle$ is the known state for the unperturbed Hamiltonian
(that is, for instance, the weave state $|\Delta\rangle$), and hence
satisfies $\mathcal{H}_0 |\psi_0\rangle=0$ and $\langle\psi_0|{\hat{q}}%
_{ab}|\psi_0\rangle=\delta_{ab}+\mathcal{O}(\ell _{P})$, while $%
|\psi_2\rangle$ is a correction due to the electromagnetic field.\newline

Now, in order that the peakedness condition stated above be satisfied, the
perturbation $|\psi_2\rangle$ must be such that
\begin{equation}
2 Re(\langle\psi_0|{\hat{q}}_{ab}|\psi_2\rangle)= \gamma_{ab}
\end{equation}
modulo corrections of the order of $\ell _{P}$.

In the following we will assume that it is possible to find a state $|\psi
_{2}\rangle $ that satisfies this expression and, hence, approximates the
classical metric derived in the previous subsection, up to corrections of
order $\ell _{P}$. Hence, we are within the assumptions made to derive eq. (%
\ref{field_eq}), which allows us to apply the formalism developed in section
3.

\subsection{The field equations: semiclassical photon propagation with back
reaction}

Recall from section 3 that the field equations are given by

\begin{eqnarray}  \label{field}
\partial_t A&=&-E_a, \\
\partial_t(H^{ab}E_b)&=&H^{ab}e_b{}^{cd}\nabla_cB_d.  \notag
\end{eqnarray}

where the metric operator $H^{ab}$ associated to the classical 3-metric $%
q_{ab}$ is

\begin{equation}
H^{ab}=\sqrt{det(q)}\left(q^{ab}+2\xi\sqrt{det(q)}\;e^{abc}\nabla_c\right).
\end{equation}

In the case under consideration, the metric and the electromagnetic field
are both given as perturbative expansions, i.e., keeping only the first
non-trivial correction

\begin{eqnarray}  \label{exp_F}
q_{ab} &=&\delta _{ab}+\epsilon ^{2}\gamma _{ab}, \\
F_{\mu\nu} &=&\epsilon F_{\mu\nu}^{(1)}+\epsilon ^{3}F_{\mu\nu}^{(2)},
\end{eqnarray}%
and therefore, the metric operator $H^{ab}$ is also given in perturbative way

\begin{equation}
H^{ab}=H^{(0)ab}+\epsilon ^{2}H^{(2)ab}
\end{equation}%
with

\begin{eqnarray}  \label{metric_0}
H^{(0)ab}&=&\delta^{ab}+2\xi \epsilon^{abc}\partial_c, \\
H^{(2)ab}&=& \left(\frac{1}{2}\delta^{ab}\gamma-\gamma^{ab}\right)+2\xi
\left((\tilde{e}^{abc}+\gamma\epsilon^{abc})\partial_c+\epsilon^{abc}%
\nabla^{(2)}_c\right).
\end{eqnarray}

In these expressions $\gamma$ is the trace of $\gamma_{ab}$, i.e., $%
\gamma=\delta^{ab}\gamma_{ab}$, and $\tilde{e}^{abc}$ and $\nabla^{(2)}_a$
are the first non trivial corrections to $e^{abc}$ and the covariant
derivative respectively,

\begin{equation}
\tilde{e}^{abc}=\frac{1}{2}\gamma \epsilon ^{abc}-3\epsilon ^{\lbrack
ab}{}_{d}\;\gamma ^{c]d},
\end{equation}%
while $\nabla _{a}^{(2)}$ applied to a co-vector $C_{b}$ is given by

\begin{equation}
\nabla^{(2)}_a C_b=-\frac{1}{2}(\partial_a \gamma_{bd}+ \partial_b
\gamma_{ad}-\partial_d \gamma_{ab})\delta^{cd}C_c.
\end{equation}

Notice that the zeroth order of the metric operator (eq. (\ref{metric_0}))
is, as expected, just the flat operator obtained in previous works \cite%
{KoPa, GleiKoPa}. \newline
\newline
On the other hand, the perturbative expansion (\ref{exp_F}) gives rise to
similar expansions for the electric and magnetic fields (see eqs. (\ref%
{pert_E}) and (\ref{pert_B})) which, inserted in the field equations (\ref%
{field}) lead to

\begin{eqnarray}
\partial _{t}\vec{A}^{(1)} &=&-\vec{E}^{(1)}, \label{firts_order1}\\
\partial _{t}(\vec{E}^{(1)}+2\xi \nabla \times \vec{E}^{(1)}) &=&\nabla
\times (\vec{B}^{(1)}+2\xi \nabla \times
\vec{B}^{(1)}),\label{firts_order2}
\end{eqnarray}%
for the first order, corresponding to the free Maxwell field propagating in
a flat background (here we have adopted vectorial notation for simplicity),
and

\begin{eqnarray}
\partial _{t}A_{a}^{(3)} &=&-E_{a}^{(3)},  \notag \\
\partial _{t}(H^{(0)ab}E_{b}^{(3)})-H^{(0)ab}\epsilon _{b}{}^{cd}\partial
_{c}B_{d}^{(3)} &=&-[\partial _{t}(H^{(0)ab}E_{b}^{(1)})-H^{(2)ab}\epsilon
_{b}{}^{cd}\partial _{c}B_{d}^{(1)}]  \notag \\
&&+H^{(0)ab}[\tilde{e}_{b}{}^{cd}\partial _{c}+\epsilon _{b}{}^{cd}\nabla
_{c}^{(2)}]B_{d}^{(1)},  \notag
\end{eqnarray}%
for the correction generated by back reaction effects. The equations for the
free field (\ref{firts_order1}) and (\ref{firts_order2}) coincide, of
course, with the ones obtained in \cite{KoPa} and \cite{GleiKoPa}, and
preserve Lorentz invariance, while the above corrections will break that
symmetry. The easiest way to see this is by considering the wave-like
equation to be satisfied by the potential (eq. (\ref{general_wave})), where
it becomes clear which term breaks covariance. From this expression we can
see that Lorentz Invariance will be broken whenever the time derivative of $%
\sqrt{det(q)}$ is different form zero. In our perturbative approach this
quantity is given by

\begin{equation}
\sqrt{det(q)}=1+\frac{1}{2}\;\epsilon ^{2}\gamma ,
\end{equation}%
whose time derivative is in general non vanishing. Hence, even for a flat
background, if we take into account back reaction effects on the metric, the
resulting propagation equations for the electromagnetic field will break
Lorentz Invariance.

Just for completeness, this wave equation in the case of interest reads, for
the free Maxwell field

\begin{equation}
\square A_{b}^{(1)}=0,
\end{equation}

and, for the back reaction correction,

\begin{eqnarray}
\lbrack (1-2\xi \nabla \times )\square A^{(3)}]^{a} &=&-\partial
_{t}(H^{(2)ab}E_{b}^{(1)})  \notag \\
&&+\left( \epsilon _{b}{}^{cd}(H^{(0)ab}\nabla _{c}^{(2)}+H^{(2)ab}\partial
_{c})+\tilde{e}_{b}{}^{cd}H^{(0)ab}\partial _{c}\right) B_{d}^{(1)}.  \notag
\end{eqnarray}

\section{Summary and conclusions}

We have studied the propagation of light in two different scenarios

\begin{enumerate}
\item On an arbitrary quantum curved spacetime in the semiclassical
approximation of Loop Quantum Gravity.

\item On a deviation from the quantum flat metric were the non trivial part
is the back reaction effect of the Maxwell field.
\end{enumerate}

For the first part we obtained the effective interaction Hamiltonian for the
gravitational and electromagnetic fields and derived the corresponding field
equations, which can be combined to obtain a wave like equation for the
Maxwell potential.\newline

In the particular case of a flat background, this wave equation reduces to
the usual Lorentz invariant propagation. This result is also valid for any
stationary background geometry, but it is no longer true for the general
case, in which the wave equation contains a covariance breaking term that is
related to the time derivative of the determinant of the 3-dimensional
metric.\newline

As an example of this we studied light propagation in flat FRW cosmology
dominated by matter, and solved the wave equation to obtain an effect that
in principle can be observed, namely, that the polarization direction of an
initial linearly polarized plane wave rotates with a frequency dependent
angle. However, it is not clear if the assumptions made to obtain these
results were too restrictive. It could happen that the assumptions cannot be
maintained for the time of flight of the photons and thus there are no
physical predictions to be made. On the other hand, if the set of
assumptions are valid there are observational consequences, such us the loss
of polarization of a linearly polarized wave packet with a frequency
spectrum. Note also that the polarization direction has a linear dependence
on the photon energy, a different result from that obtained by Gambini and
Pullin where the dependence is quadratic \cite{GaPu}. However, the fact that
we do observe light with a large amount of linear polarization tells us
that, if this effect actually exists, it must be much smaller than expected,
and, moreover, using recent observational data it is possible to put a very
severe bound on the phenomenological constant $\xi $ \cite{GleiKo}.\newline

A second and for us more important problem was to analyze the propagation of
light taking into account back reaction effects. We have seen that Lorentz
invariance is also broken and, although we have not solved the equations, it
is reasonable to believe that wave propagation would present similar effects
to the ones obtained for a FRW background. However, in this case, any
induced effect, if present, would be much more difficult to observe since it
is of higher order: the corrections are second order in the small parameter $%
\epsilon $ and, besides, of order $\xi \ell _{p}$.\newline

It is worth mentioning that polarized light travelling on a media with an
index of refraction induced by the quantum spacetime is a very sensitive
tool to study these corrections and it is a worthwhile problem to obtain a
predicted value for the rotation of the polarization direction.

As a final comment we would like to mention that it is surprising to
obtain noncovariant semiclassical equations of motion coming from a
covariant formalism. Loop quantum gravity is by construction a
covariant theory, although the 3+1 splitting hides this fact. The
only possible place where this covariance can be broken is in the
use of semiclassical states. These states are only gauge invariant
with respect to the rotation group on the spacial surface but do not
satisfy the hamiltonian constraint (otherwise they cannot be peaked
around a classical metric). Maybe it is impossible to define a
covariant semiclassical approximation of loop quantum but this would
be a rather unwanted feature on the theory.


\end{document}